\documentclass[
               showpacs,            
               preprintnumbers,     
               aps,                 
               prd,                 
               a4paper,             
               superscriptaddress,  
               nofootinbib,         
               tightenlines,        
               floats,floatfix      
               ]{revtex4}
\usepackage{graphicx}
\usepackage{latexsym}
\usepackage{
amssymb}        

\renewcommand\[{\left[}
\renewcommand\]{\right]}

\newcommand\eq[1]{Eq.~(\ref{#1})}

\newcommand\eqref[1]{(\ref{#1})}

\newcommand\ee{\end{equation}}
\newcommand\be{\begin{equation}}
\newcommand\eea{\end{eqnarray}}
\newcommand\bea{\begin{eqnarray}}

\def\cala{{\cal A}}

\def\calp{{\cal P}}
\def\calr{{\cal R}}

\newcommand\bfk{{\bf k}}

\newcommand\bfx{{\bf x}}

\newcommand\sub[1]{_{\rm #1}}

\newcommand{\vev}[1]{{\langle #1 \rangle }}

\begin{document}

\title{
Forming sub-horizon black holes at the end of inflation}
\author{David H.~Lyth}
\affiliation{Physics Department, University of Lancaster, Lancaster LA1 4YB, 
United Kingdom}
\author{Karim A.~Malik}
\affiliation{Physics Department, University of Lancaster, Lancaster LA1 4YB, 
United
Kingdom}
\author{Misao Sasaki}
\affiliation{Yukawa Institute for Theoretical Physics, Kyoto University, Kyoto 
606-8502, Japan}
\author{Ignacio Zaballa}
\affiliation{Physics Department, University of Lancaster, Lancaster LA1 4YB, 
United Kingdom}
\date{\today}
\pacs{04.70.-s, 98.80.-k \hfill astro-ph/0510nnn}

\begin{abstract}
Previous authors have calculated the mass function of primordial black
holes only on scales which are well outside the horizon at the end of
inflation.  Here we extend the calculation to sub-horizon scales, on
which the density perturbation never becomes classical. Regarding the
formation of black holes as a `measurement' of the (high peaks) of the
density perturbation, we estimate a mass function by assuming that
black holes form as soon as inflation ends, in those rare regions
where the Bardeen potential exceeds a threshold value of 
$\Psi_{\rm c}\simeq0.5$.
\end{abstract}

\maketitle

\section{Introduction}

During inflation the vacuum fluctuation of each light field is
promoted around the time of horizon exit to a classical perturbation.
According to present thinking, one or more of these light field
perturbations is responsible for all primordial perturbations in the
Universe~\cite{Liddle:2000cg}. It is
usually supposed that primordial perturbations are absent on those
very small scales which are still inside the horizon at the end of
inflation, on the ground that the vacuum fluctuation on such scales
never becomes classical.

These ideas are usually applied to the perturbation on cosmological
scales, which at least in the standard cosmology are far outside the
horizon by the end of inflation.  
A different application though is to the formation of Primordial Black
Holes (PBH)~\cite{Carr:1974nx,Carr:1975qj}. After inflation, as each
scale comes back inside the horizon, an appreciable number of black
holes (BH) will form provided that the mean-square energy density
contrast smoothed on that scale is not too far below unity. The black
holes form in those relatively rare regions where the density contrast
is actually of order unity, and their mass is of order the total energy
inside the horizon. Assuming a specific form for the spectrum of the
density contrast, this allows one to estimate the mass function of the
black holes using for instance the Press-Schechter formalism~\cite{PS}
\footnote{
The original PBH abundance calculations~\cite{Carr:1974nx,Carr:1975qj}
and early numerical studies, e.g.~\cite{Niemeyer:1999ak}, used
the density contrast (on comoving velocity orthogonal slicings) to
study the gravitational collapse.
In a recent study~\cite{Green:2004wb} (see also \cite{Shibata:1999zs})
a metric perturbation, the curvature
perturbation on uniform density hypersurfaces was used instead, giving
excellent agreement with detailed numerical 
investigations~\cite{Musco:2004ak}.}.
PBH abundance calculations provide a powerful tool to study
inflationary models through their primordial power spectrum over a
wide range of scales which are difficult to constrain by other data
from Large Scale Structure surveys or Cosmic Microwave Background
experiments~\cite{Carr:1994ar,Kim:1996hr,Green:1997sz,Leach:2000ea}.
For a comprehensive recent review on PBHs see e.g.~\cite{Carr:2005bd}.

The standard PBH calculation makes sense on those scales which are
well outside the horizon by the end of inflation, but what is supposed
to happen on smaller scales? As the comoving wavenumber $k$ is
increased 
above that wavenumber $k\sub{end}$ which leaves the
horizon at the end of inflation, the standard calculation ceases to
make sense and a different calculation must take its place, which
takes on board the essentially quantum nature of the vacuum
fluctuation.  In this paper we explain for the first time how such a
calculation should proceed.

The layout of the paper is as follows: in the next section we briefly
review how fluctuations in the inflaton field are generated during
inflation and calculate the spectrum of the fluctuations at the end of
inflation.  In Section \ref{pert_sec} we use the primordial power
spectrum as input for the Bardeen potential and calculate how the
latter then evolves after the end of inflation during radiation
domination. In Section \ref{subhorizon_sec} we investigate how black
holes are formed on sub-horizon scales using the Press-Schechter
formalism and calculate the mass function of these black holes.  We
discuss our results and conclude in Section \ref{disc_sec}.

\section{The vacuum fluctuation}
\label{vacuum_sec}

During inflation the vacuum fluctuation of each light (practically
massless) field is promoted to a classical perturbation, on those
scales which leave the horizon. We shall be interested in
single-component inflation, during which the only relevant light field
is the slowly-rolling inflaton field $\phi$. The perturbation of the
field, $\delta\phi$, is related to the curvature perturbation ${\cal {R}}$
on comoving, or uniform field hypersurfaces, by
\be
\label{defR}
{\cal {R}}= -\frac{H}{\dot\phi} \delta\phi \,,
\ee
where the right hand side of \eq{defR} has to be evaluated on flat slices. 
In the leading order of the slow-roll approximation the field equation for
$\delta\phi$ on flat slices
lives in unperturbed spacetime. (In other words, we can
ignore the back-reaction of the metric perturbation.)

In this section we recall the standard description of the behaviour
during inflation of the vacuum fluctuation of $\phi$, which actually
applies to the vacuum fluctuation of any practically massless field.
We use conformal time $\tau$ to work with $\varphi\equiv a\delta\phi$, where
$a$ is the scale factor. We take inflation to be exponential
corresponding to constant Hubble parameter $H$, and take
\be
\tau = -(aH)^{-1} \,,
\ee
so that $\tau=0$ corresponds to the infinite future.

We consider the Fourier component of $\varphi$ with comoving
wave vector $\bfk$, whose classical field equation is
\be
\frac{d^2 \varphi(\bfk,\tau)}{d\tau^2} +\left(k^{2}-\frac{2}{\tau^2}
     \right)\varphi(\bfk,\tau)=0
\,.
\label{ueom}
\ee
Well before horizon exit \eq{ueom} reduces to
\be
\frac{d^2 \varphi(\bfk,\tau)}{d\tau^2} + k^2  \varphi(\bfk,\tau) = 0
\,.
\label{ueom2}
\ee
This is the equation of a unit-mass harmonic oscillator with angular
frequency $k$.

Now we quantise the perturbation $\delta\varphi$ while keeping the
unperturbed field classical.  At the quantum level, each observable
corresponds to a hermitian operator.  We use a hat to denote
operators, so that the operator corresponding to
$\delta\varphi(\bfx,\tau)$ is $\widehat{\delta\varphi}(\bfx,\tau)$.
The `reality condition' ensuring that
$\widehat{\delta\varphi}(\bfx,\tau)$ is hermitian is
\be
\hat\varphi(-\bfk,\tau) =
\hat\varphi^\dagger(\bfk,\tau)
\,,
\label{x}
\ee
where the dagger denotes the hermitian conjugate.

As is usual in quantum field theory, we adopt the Heisenberg picture,
in which the state vector describing the system is time-independent
and the operators corresponding to observables have the classical
time-dependence. Then $\hat\varphi(\bfk,\tau)$ is a solution of
\eq{ueom}, and to satisfy the reality condition it must be of the form
\be
\label{eq:reality}
\hat\varphi(\bfk,\tau) = \varphi(k,\tau) \hat a(\bfk) +
\varphi^*(k,\tau) \hat a^\dagger(-\bfk),
\label{reality}
\ee
where $\varphi(k,\tau)$ is a solution of \eq{ueom}, and $\hat a$ is a
time-independent operator.

To define $\varphi(k,\tau)$ and $\hat a$, we focus on the era well
before horizon exit,
and consider a region of space-time much smaller
than $H^{-1}$ (i.e.\ with $\Delta \tau$ and $\Delta x$ both much less
than $(aH)^{-1}$).  In such a region space-time is practically flat
(in other words, gravity is negligible) and we can assume that flat
space-time field theory applies.  The scale factor $a$ is practically
constant, and we can make the choice $a=1$ instead of the usual
$a_0=1$. Then $\bfk$ is the physical wave-vector, and $\varphi$ is the
original field $\phi$.

Flat space-time field theory is constructed in such a way as to arrive
at the particle concept. It boils down for scalar fields to noticing
that each Fourier component has the dynamics of a unit-mass harmonic
oscillator with position $q=\varphi(\bfk)$.  The vacuum corresponds to
all of the oscillators being in the ground state, and the
equally-spaced excited energy levels of each oscillator correspond to
the presence of $1$, $2,\cdots$ massless particles, each with momentum
$\bfk$.  All of this is encoded by the flat space-time field theory
mode function
\be
\varphi(k,\tau)= \frac1{\sqrt{2k}} e^{-ik\tau}
\label{896}
\label{894}
\,,
\ee
the commutation relations
\be
[\hat  a(\bfk),\hat a^\dagger(\bfk')]=
 \delta^3 (\bfk-\bfk')
 \qquad
[\hat  a(\bfk),\hat a(\bfk')]= 0
\label{com}
\,,
\ee
and the assertion that $\hat a$ annihilates the vacuum state,
\be
\hat a(\bfk) | {\rm vacuum} \rangle =0
\label{annih}
\,.
\ee

The solution of \eq{ueom} satisfying the initial condition \eqref{896}
is 
\be
\varphi(k,\tau)=  \frac{1}{\sqrt{2k}} \, \left(1 + i{k\tau}
\right) \frac{1}{ik\tau}e^{-ik\tau}
\label{masslessmode}
\,.
\ee

We are supposing that the time-independent state vector of the
Universe corresponds to the vacuum. At any instant, the vacuum is a
superposition of states with definite values for the Fourier
components $\varphi(\bfk)$. Consider now some particular instant, and
imagine that the Fourier components are measured. As the Fourier
components are uncorrelated, the perturbation is gaussian so that all
of its stochastic properties are determined by the two-point
correlator,
\be
\langle \hat\varphi(\bfk) \hat\varphi(\bfk') \rangle
= |\varphi^2(k)|^2 \delta^3(\bfk+\bfk') \,.
\ee

The spectrum of $\phi$ is defined as
\be
\calp_\phi= \frac {k^3}{2\pi^2a^2} |\varphi^2 (k) |
\,,
\label{calpu98}
\label{phispec}
\ee
the normalisation chosen so that the mean square is
\be
\vev {|\delta\phi|^2 } = \int^\infty_0 \calp_\phi(k)  \frac{dk}{k}
\,.
\ee
Well after horizon exit it reduces to the famous result $\calp_\phi =
(H/2\pi)^2$.

The previous paragraph holds equally before and after horizon exit.
The issue of quantum versus classical arises when we consider the
evolution after the measurement has been made.  The perturbation
$\hat\varphi(\bfk)$ can be regarded as classical if there exist states,
analogous to the ``wave packets'' that one considers in ordinary
quantum mechanics, which are practically eigenstates of
$\hat\varphi(\bfk)$ over a long period of time. As we are working in
the Heisenberg picture, this means that the perturbation is classical
if and only if its operator has almost trivial time-dependence,
\be
\hat \varphi(\bfk,\tau) \simeq f(\tau)\hat\varphi(\bfk,\tau_0)
\,,
\ee
where $f$ is a number and $\tau_0$ is some initial time. Such is the
case well after horizon exit, when~\cite{Starobinsky82,Starobinsky86}
\be
\hat \varphi(\bfk,\tau) \simeq   \frac i{\sqrt{2k^3}\tau}
\[ \hat a(\bfk) -  \hat a^\dagger(-\bfk) \]
\,,
\ee
but not earlier.

This is the usual justification for regarding the field perturbation
as classical well after horizon exit.  It is not completely
satisfactory since in the context of cosmology one can hardly talk
about a complete measurement, but it works in practise.  
(Some of the conceptual issues are addressed in the extensive
literature on decoherence, e.g.~\cite{decoherence}, and on the
quantum-to-classical transition~\cite{Polarski:1995jg,Kiefer:1998qe}.)
Our concern here is with those Fourier components $\hat\varphi(\bfk)$,
which at the end of inflation are still inside the horizon or at least
not very far outside it.

Limiting ourselves to the issue of black hole formation, we propose
that the formation of the black holes should itself be regarded as a
measurement.  Accordingly, we will go ahead and calculate the mass
function of the black holes, using the spectrum \eqref{phispec}
exactly as if it referred to a classical quantity. In the end we will
confine ourselves to the sub-horizon regime $k\gg k\sub{end}$ in order
to arrive at a simple and fairly model-independent result, but it is
clear that within the context of a given model our methods would allow
a continuous mass function to be calculated covering also the regime
$k\sim k\sub{end}$.

Two notes of caution on the applicability of our results are in order here.
Firstly, during the inflationary stage, since the amplitude of ${\cal R}$
increases as $k/aH$ increases, there is a scale below which it becomes
greater than unity. It is given by $k/aH=1/{\cal R}_{HC}$, 
where ${\cal R}_{HC}$ is the value of ${\cal R}$ at the horizon scale.
Since we do not know how to deal with this scale, we confine our arguments
to scales greater than this critical scale.
Secondly, inside the horizon, the metric perturbations due to second
order matter perturbations can in general become important. In the
case of the quantum fluctuations, this seems to give the condition
that the scale should be greater than $V'(\phi)/H^2$, or $k/aH <
V'/H^3$, where $V$ is the inflaton potential and a dash denotes
differentiation with respect to $\phi$. Below this scale, the second
order perturbations dominate.
Hence the arguments in this article apply only to scales larger than
these lower bounds.

\section{Cosmological Perturbations}
\label{pert_sec}

\subsection{The evolution of the Bardeen potential and the 
curvature perturbation}

The line element of the unperturbed spatially flat FRW universe is
defined as
\be
ds^2=a^2(\tau)(-d\tau^2+\delta_{ij}dx^idx^j)\,,
\ee
where $a(\tau)$ is the scale factor, $\tau$ is the conformal time,
$\delta_{ij}$ is the metric of the flat background, and $\mathbf{x}$
are the spatial coordinates. The perturbed metric in the longitudinal
gauge, considering only scalar perturbations, is given by
\be
ds^2=a^2\left(\tau\right)\left[
-\left(1+2\Psi\right)d\tau^2+\left(1+2\Phi\right)\delta_{ij}dx^idx^j
\right]  \,.
\ee  
In this gauge the metric perturbations coincide with the gauge
invariant Bardeen potentials defined in \cite{Bardeen:1980kt},
$\Psi=\Phi_A$ and $\Phi=\Phi_H$. $\Psi$ is the lapse function that
determines the proper time intervals between zero shear hypersurfaces
(the slicing of the longitudinal gauge). $\Phi$ is the curvature
perturbation on slices of zero shear~\cite{Kodama:1985bj}.

The relation of these two potentials to the density perturbation is
very insightful. In the cases of a scalar field and a perfect fluid
there is no anisotropic stress, and then $\Phi=-\Psi$. 
The potential $\Psi$ is then related the density perturbation in the
comoving gauge~\cite{Liddle:2000cg}\footnote
{Here and in the following, quantities with specified wavenumber $k$ are
 mode functions governing the evolution of the corresponding 
operators. Thus $\delta_k$ for instance is defined through \ref{eq:reality}, with 
$\hat \varphi(\bfk,\tau)$ replaced by $\hat \delta(\bfk,\tau)$ and
$\varphi(k,\tau)$ replaced by $\delta(k,\tau)\equiv \delta_k$.}
\begin{equation}
\label{eq:Poissoneq}
\delta_{k}\equiv\frac{\delta\rho_k}{\rho}
=-\frac{2}{3}\left(\frac{k}{aH}\right)^2\Psi_k\,,
\end{equation}
where the perturbation $\delta_k$ is the density contrast defined in
the comoving slicing, $\rho$ is the density of the unperturbed
background, and $H$ is the corresponding Hubble parameter, $H^2=8\pi
G\rho/3$. Expression (\ref{eq:Poissoneq}) is analogous to the Poisson
equation in Newtonian gravity.

The Fourier components of $\cal{R}$, defined in \eq{defR} above, are
simply related to the Fourier components of the field fluctuations on
flat slices $\delta\phi$ by
\be
\label{eq:curvatureR}
{\cal{R}}_{k}=-\frac{H}{\dot\phi} \delta\phi_{k}\,.
\ee
The curvature perturbation ${\cal{R}}_k$ sets to a constant value a
few Hubble times after horizon crossing, getting a very small scale
dependence determined by the value that $H$ and $\dot\phi$ have at
that time. 
Note that this result is obtained by considering the evolution
equation for ${\cal R}_k$, but not from the approximate solution
for $\varphi_k=a\delta\phi_k$ given by Eq.~(\ref{masslessmode}). 
 On scales inside the horizon, the time dependence of
${\cal{R}}_k$ is mainly given by the vacuum fluctuations in
\eqref{masslessmode}. Then one may write
\bea
\label{eq:Rinfl}
{\cal{R}}_{\rm{k}}\left(t\right)= \left\{ \begin{array}{ll}
-\frac{1}{\sqrt{2k^3}}\frac{H^2}{\dot\phi}\left(i+\frac{k}{aH}\right){\rm{exp}}
\left(\frac{ik}{aH}\right) & \textrm{for $k\gtrsim aH$}\,,\\
\\
-\frac{i}{\sqrt{2k^3}}\left(\frac{H^2}{\dot\phi}\right)_{{t=t}_*} & 
\textrm{for $k\ll aH$}\,,
\end{array}\right.
\eea
where ${t}_*$ indicates the time when a given scale $k$ leaves the
horizon.

The other quantity we are interested to determine from the vacuum
fluctuations is the Bardeen potential $\Psi$. In the absence of
anisotropic stress the evolution of $\Psi$ is given by the following
first order equation
\begin{equation}
\label{eq:Rconservation}
\frac{2}{3}H^{-1}\dot{\Psi}_k+\frac{\left(5+3w\right)}{3}\Psi_k
=-\left(1+w\right){\cal{R}}_k\,.
\end{equation}
where $w=p/\rho$. On super-horizon scales $\mathcal{R}_k$ is constant,
and the decaying mode can be neglected. Assuming $w$ is only weekly
time-dependent, the solution is given by
\be
\label{eq:longwav}
\Psi_k=-\frac{3\left(1+w\right)}{5+3w}\mathcal{R}_k\,.
\ee
During slow-roll inflation $w\simeq-1$ and therefore
$|\Psi_k|\ll|\mathcal{R}_k|$. 
Inside the horizon $\mathcal{R}_k$ varies with time,
and the time derivative under slow-roll conditions from
(\ref{eq:Rinfl}) is given by
\begin{equation}
\dot{\cal{R}}_k
=\frac{iH^3}{\sqrt{2k^3}\dot\phi}\left(
\frac{k}{aH}\right)^2{\rm{exp}}\left(i\frac{k}{aH}\right)\,.
\end{equation}
On the other hand, $\dot\mathcal{R}_k$ can be written as
\cite{Liddle:2000cg}
\begin{equation}
\dot\mathcal{R}_k
=-H\frac{\delta P_k}{\rho\left(1+w\right)}=
\frac{2}{3}H\left(\frac{k}{aH}\right)^2\frac{\Psi_k}{\left(1+w\right)}\,,
\end{equation}
where $\delta P_k$ is the pressure perturbation in comoving
hypersurfaces, and $\delta P_k=\delta\rho_k$ for the inflaton
field. Comparing both expressions for $\dot\mathcal{R}_k$, the mode
function for the Bardeen equation yields
\begin{equation}
\label{Phi_kfinal}
\Psi_k=\frac{3i}{2}
\frac{H^2\left(1+w\right)}{\sqrt{2k^3}\dot\phi}{\rm{exp}}
\left(i\frac{k}{aH}\right)\,,
\end{equation}  
which together with the previous result on super-horizon scales,
Eq.~(\ref{eq:longwav}), shows that $\Psi_k$ during slow-roll is
practically zero on all scales, provided that $H^2/|\dot\phi|\ll1$
which is required by observational consistency.

\subsection{Junction Conditions}

In this section we are setting the initial conditions for the
radiation dominated epoch after the end of inflation. If reheating
occurs almost instantaneously, then we can use the values of for
$\calr_k$ and $\Psi_k$ at the end of inflation as the initial
conditions for the radiation epoch, as we will show below. A hybrid
inflationary model, in which the decay and thermalisation of the
particles produced during reheating takes place very rapidly, is most
appropriate to realise the definite values of these quantities. In
these models, the end of inflation is determined by the critical value
of the inflaton field $\phi_{\rm c}$, which triggers the rapid fall of the
secondary field $\sigma$ to its true vacuum. Then the matching of the
solutions must be done on comoving hypersurfaces,
$\phi_{\rm c}(\mathbf{x},t)=\rm{constant}$.

The junction conditions require the intrinsic metric and the extrinsic
curvature tensor of the comoving 3-hypersurfaces on which we match to
be continuous~\cite{Israel:1966rt,Deruelle:1995kd,Martin:1997zd}.
{}From the requirement on the intrinsic metric we therefore find that
$\calr$ has to be continuous.
The trace of the extrinsic curvature of a comoving hypersurface at a
given point $\mathbf{x}$, is the locally defined Hubble parameter
$H(\mathbf{x},t)=H(t)+\delta H(\mathbf{x},t)$. Therefore the
perturbation $\delta H_k$ is continuous, and it can be written as
\be
2H\delta H_k
=\frac{8\pi G}{3}
\delta\rho_k-\frac{2}{3}\left(\frac{k}{a}\right)^2\mathcal{R}_k\,,
\ee
showing that the Bardeen potential $\Psi_k$, related to the comoving
energy density $\delta\rho_k$ by (\ref{eq:Poissoneq}), is also
continuous.

Since ${\cal{R}}_k$ and $\Psi_k$ are continuous at the transition time
in the comoving gauge, from Eq.~(\ref{eq:Rconservation})
$\dot{\Psi}_k$ must be discontinuous on that hypersurface. Its value
can be determined at the beginning of the radiation epoch from
(\ref{eq:Rconservation}), setting $\Psi_k=0$ and taking
$\mathcal{R}_k$ given in (\ref{eq:Rinfl}), and is given by
\bea
\label{eq:phidotend}
\dot{\Psi}_k\left(t_{\rm e}\right)\vert_{\rm{rad}}= \left\{\begin{array}{ll}
-\frac{3}{2}(1+w_0)H\left(t_{\rm e}\right)\mathcal{R}_k\left(t_{\rm e}\right)&
 \textrm{for $k\gtrsim a_{\rm e}H_{\rm e}$}\,,\\
\\
-\frac{3}{2}(1+w_0)H\left(t_{\rm e}\right)\mathcal{R}_k\left(t_*\right)&
 \textrm{for $k\ll a_{\rm e}H_{\rm e}$}\,,
\end{array}\right.
\eea
where $w_0=p/\rho=c_s^2=1/3$ is the velocity of sound squared
during radiation domination, $a_{\rm e}$ and $H_{\rm e}$ are the scale factor
and the Hubble parameter at the end of inflation, respectively, and
$\calr_k(t_*)$ denotes the value of the curvature perturbation when
the scale leaves the horizon given in (\ref{eq:Rinfl}).

To smoothly match the background scale factor between the inflationary
stage and the radiation-dominated stage, it is necessary to make
a constant time shift of the conformal time $\tau$.
We set
\begin{eqnarray}
a(\tau)=\left\{
\begin{array}{ll}
\displaystyle\frac{1}{H(\tau-2\tau_{\rm e})}\quad &\mbox{for}~\tau<\tau_{\rm e}\,,
\\
~\\
\displaystyle\frac{\tau}{H\tau_{\rm e}^2}\quad&\mbox{for}~\tau\geq \tau_{\rm e}\,,
\end{array}\right.
\end{eqnarray}
where $\tau_{\rm e}$ is the conformal time at the end of inflation.
Thus, $\tau$ in all the previous formulae should be replaced
by $\tau-2\tau_{\rm e}$.

\subsection{The Bardeen potential during radiation domination}
\label{sec:radiation}

For a radiation dominated epoch the Bardeen potential is
\cite{Bardeen:1980kt}
\begin{equation}
\label{eq:Bardeenrad}
\Psi_k\left(\tau\right)=\left(\frac{1}{x}\right)^3
\biggl[a_1\left(x\cos x-\sin x\right)+a_2\left(x\sin x+\cos x\right)\biggr]\,,
\end{equation}
where $a_1$ and $a_2$ are integrating constants, and
$x=c_sk\tau$.

From $\Psi_k=0$ one gets the following relation between $a_1$ and $a_2$
\begin{equation}
\label{eq:radconditions11}
a_1=-A\left(k,\tau_{\rm e}\right)a_2\,,
\end{equation}
where $x_{\rm e}=c_sk\tau_{\rm e}$, and $A\left(k,\tau_{\rm e}\right)$ is
given by 
\begin{equation}
\label{eq:Aktau}
A\left(k,\tau_{\rm e}\right)
=\frac{\left(x_{\rm e}\sin x_{\rm e}+\cos x_{\rm e}\right)}{\left(x_{\rm e}\cos x_{\rm e}-\sin x_{\rm e}\right)}.
\end{equation}
Differentiating and evaluating Eq.~(\ref{eq:Bardeenrad}) at
$\tau_{\rm e}=1/a_{\rm e}H_{\rm e}$ yields
\begin{equation}\label{eq:Bardeenraddiff}
\frac{d\Psi_k}{d\tau}\left(\tau_{\rm e}\right)
=\frac{a_2}{c_sk\tau_{e}^2}\biggl[\cos x_{\rm e}+A\left(k,\tau_{\rm e}\right)\sin x_{\rm e}
\biggr]\,.
\end{equation}
Then taking the derivative of Eq.~(\ref{eq:phidotend}), the
integrating constants are
\begin{eqnarray}
\label{eq:a_2}
a_2&=&c_sk\tau_{e}^2\biggl[
-\frac{3}{2}\left(1+w\right)\frac{{\cal{R}}_k}{\tau_{\rm e}}
\biggr]\biggl[\cos x_{\rm e}+A\left(k,\tau_{\rm e}\right)\sin x_{\rm e}\biggr]\nonumber\\
&=&-\frac{3}{2}\left(1+w\right){\cal{R}}_k\left(x_{\rm e}\cos x_{\rm e}-\sin x_{\rm e}
\right)\,,\\
a_1&=&\frac{3}{2}\left(1+w\right){\cal{R}}_k\left(x_{\rm e}\sin x_{\rm e}+\cos x_{\rm e}\right)
\label{eq:a_1}\,.
\end{eqnarray}
Substituting the value of $a_1$ and $a_2$ into
Eq.~(\ref{eq:Bardeenrad}), one finds that $\Psi_k$ during radiation
domination is
\be
\label{eq:radsolution}
\Psi_k\left(\tau\right)
=\frac{3\left(1+w_0\right)\mathcal{R}_k}{2x^3}\biggl[
\left(x-x_{\rm{e}}\right)\cos(x-x_{\rm e})-\left(1+xx_{\rm{e}}\right)\sin(x-x_{\rm e})
\biggr]\,.
\ee
The value of $\calr_k$ in this last expression is given by
Eq.~(\ref{eq:Rinfl}) evaluated at $\tau_{\rm e}$. On scales $k\ll a_{\rm e}H_{\rm e}$,
$\sqrt{k^3}\,\calr_k\sim (H^2/\dot\phi)_{\rm{\tau=\tau}_*}$, 
and it has a slight scale dependence, while for $k\gtrsim a_{\rm e}H_{\rm e}$ 
it has the same fixed value for all scales inside the horizon at $\tau_{\rm e}$.

Long after the end of inflation, $\tau\gg\tau_{\rm e}$, and on scales
bigger than the sound horizon, $x=kc_{\rm s}\tau<1$, we recover the
usual long wavelength result for $\Psi_k$, given by
\eq{eq:longwav}. Using $w_0=1/3$ and expanding the sine and cosine
functions, we get from \eq{eq:radsolution}
\begin{eqnarray}
\label{eq:Bardeenradlong}
\Psi_k\left(\tau\right)
&=&-\frac{2\mathcal{R}_k}{x^3}\left[
\frac{x^3}{3}+x^2x_{\rm e}-O\left(x^5\right)\right]\nonumber\\
&=&-\frac{2}{3}\mathcal{R}_k\left[1+\frac{\tau_{\rm e}}{\tau}
+O\left(x^2\right)\right]\,.
\end{eqnarray}
%
which agrees to first
order in $x$ with the value obtained from Eq.~(\ref{eq:longwav}) setting
$w=1/3$.

\section{Black hole formation inside the horizon at the end of inflation}
\label{subhorizon_sec}
%
\subsection{Motivation}
\label{sec:motivationsubhorizonsize}

The solution of the Bardeen potential for radiation domination,
\eq{eq:radsolution}, shows that the oscillations of the potential are
damped with time, and the maximum value that $\Psi_{\rm{k}}(\tau)$ can
reach in the linear regime is during the first oscillation. Previous
studies indicate that the criterion for the BH formation from
super-horizon scale perturbations is $\Psi_{\rm{k}}(\tau)={\rm O}(1)$ at
horizon re-entry.  In fact, we expect this criterion to hold also for
sub-horizon scale perturbations. The reason is as follows. Consider a
perturbation localised on a physical length scale
$L$. Eq.~(\ref{eq:Poissoneq}) indicates
\begin{eqnarray}
\Psi\sim G\,\delta\rho\, L^2\sim \frac{G\,\delta M}{L}\,,
\label{Psicrit}
\end{eqnarray}
where $\delta M$ is the total mass excess of the perturbation.  On
scales sufficiently smaller than the horizon the Hubble expansion can
be neglected, and the condition $2\,G\,\delta M\gtrsim L$ is equivalent
to the usual condition for the BH formation in the asymptotically flat
spacetime. Therefore on all scales inside the horizon one expects that
the condition $\Psi\simeq1/2$ ensures that the perturbations have
entered the relativistically nonlinear regime and started collapsing. 
On the other hand from  Eq.~(\ref{eq:Poissoneq}), $\Psi_{\rm c}=0.5$ 
corresponds to a value of the critical density contrast at horizon crossing of
 $\delta_{\rm c}=1/3$, and this value of $\delta_{\rm c}$ is slightly
favoured from superhorizon PBH collapse numerical 
estimations~\cite{Green:2004wb,Shibata:1999zs,Musco:2004ak}.

One might wonder, on scales well inside the horizon if the usual
approach, considering Jeans stability criterion for the perturbations
in the linear regime, precluded black hole formation. However using
the same criterion as above for horizon scale perturbations reveals
that for $\Psi\simeq1/2$ that would not be the case.

\subsection{Mass Spectrum Estimation}
\label{subsec:massspectrumcalc}

The derivation of the mass spectrum of PBHs on scales smaller than the
horizon at the end of inflation in Press-Schechter theory resembles
the standard one where perturbations reenter the horizon after the end
of inflation~\cite{Carr:1975qj,Green:2004wb}. Here we use the
threshold criterion for PBH formation $\Psi_{\rm{c}}=0.5$, and
therefore to apply the Press-Schechter formalism we use the Bardeen
Potential instead of the density field.

The fraction of space of PBHs with mass larger than $m$ for a Gaussian
probability distribution is given by
\begin{equation}
F\left(m,\Psi_{\rm c}\right)
=\int_{\Psi_{\rm c}}^{\infty}
\frac{1}{\sqrt{2\pi}\sigma_{\Psi}\left(m\right)}{\rm{exp}}
\left(-\frac{\Psi^2}{2\sigma^2_{\Psi}\left(m\right)}\right)d\Psi \,,
\label{Massfrac}
\end{equation} 
where $\sigma_{\rm{\Psi}}(m)$ is the mean square deviation of the
Bardeen potential after smoothing the field on a given mass $m$. Using
a top hat window function in Fourier space, the mean square deviation
is approximately given by
\begin{equation}
\label{eq:meansquarephi2}
\sigma^2_{\rm{\Psi}}\left(m\right)\simeq\mathcal P_{\rm{\Psi}}
\left(m\right)\,.
\end{equation}

On sub-horizon scales the oscillating power spectrum
Eq.~(\ref{eq:radsolution}) is damped out with time. However, the
$\Psi_{\rm{k}}\sim1$ condition corresponds to strong gravitational
effects and therefore to nonlinear evolution, and as a consequence,
once a perturbation on a given scale reaches that value it evolves
nonlinearly. For that reason, to estimate the number of PBHs formed on
a certain scale at the end of inflation using Press-Schechter theory,
one needs the mean square deviation of the smoothed field at
approximately the time when the oscillating potential $\Psi_{\rm{k}}$
reaches the first maximum.
The time dependence in this calculation is handled in a similar way as
in the standard scenario. In that case as a perturbation enters the
horizon black holes are formed if the amplitude of the perturbation is
larger than the threshold value, and the size of the BH is determined
by the time (and the horizon size at that time) at which it is
formed. Here the size of a PBH is determined by the physical
wavelength at which $\Psi$ reaches the threshold value.

In the short wavelength limit, $x_{\rm e}\gg 1$, the Bardeen potential in
Eq.~(\ref{eq:radsolution}) can be written as
\be
\Psi_{\rm{k}}\left(\tau\right)
\simeq\frac{2H_{\rm{e}}^2}{\sqrt{2k^3}\dot{\phi}}
\left(\frac{x_{\rm e}^2}{c_sx^2}\right)\sin\left(x-x_{\rm{e}}\right)
\exp\left(\frac{ix_{\rm e}}{c_s}\right)\,.
\ee
In this limit, the Bardeen potential $\Psi_{\rm{k}}$ reaches the first
maximum at a time $\tau_{*}$, such that
\be
\left(x-x_{\rm e}\right)=kc_s(\tau_*-\tau_{\rm e})=\frac{\pi}{2}.
\ee
As $k$ approaches the value of horizon at the end of inflation
$k=a_{\rm e}H_{\rm e}$, the maximum of the Bardeen potential cannot be calculated
analytically since it involves a transcendental
equation. Nevertheless, it can be shown that in that limit,
$(x_*-x_{\rm e})={\rm{O}}(1)$.\footnote{The difference
$(x_*-x_{\rm{e}})$ approaches very rapidly to $\pi/2$ as $k$
increases. On scales between the causal and the sound horizon at the
end of inflation, $(x_*-x_{\rm e})<1$, while for smaller scales
$1<(x_*-x_{\rm e})<\pi/2$.}

The time at which the $\Psi_k$ potential reaches the first maximum
determines the mass of the PBH, $m$, and the smoothing scale for the
field. The mass of the PBH is given then by
\be
\label{eq:PBHmass}
m=\frac{4\pi}{3}\rho_*\left(\frac{a_*}{k}\right)^3
\simeq\frac{4\pi}{3}\rho_{\rm e}\left(\frac{a_{\rm e}}{k}\right)^3\frac{x_{\rm e}}{x_{\rm e}+1}
\,,
\ee
where $\rho_*$ is the energy density of the unperturbed background at
$\tau=\tau_*$, and $(a_*/a_{\rm{e}})^{1/2}\simeq1+ 1/x_{\rm e}$.

To estimate the mean square deviation at the maximum, first consider
the time dependent expression smoothed on the mass scale $m$, which
from Eqs.~(\ref{eq:radsolution}) and (\ref{eq:meansquarephi2}) is
given by
\begin{equation}
\label{eq:meansquarephi3}
\sigma^2_{\rm{\Psi}}\left(m,t\right)
=\frac{4{\cal{P}}_{\cal{R}}\left(x_{\rm{e}}\right)}{x^6}
\biggl[\left(x-x_{\rm{e}}\right)\cos(x-x_{\rm e})
-\left(1+xx_{\rm{e}}\right)\sin(x-x_{\rm e})\biggr]^2\,,
\end{equation}
where $\mathcal{P}_{\cal{R}}(x_{\rm e})$ is the power spectrum of the
curvature perturbation, defined by
$\mathcal{P}_{\mathcal{R}}(k,\tau)
=k^3/2\pi^2<\vert\mathcal{R}_k(\tau)\vert^2>$,
evaluated at the end of inflation. From Eq.~(\ref{eq:Rinfl}),
$\mathcal{P}_{\cal{R}}(x_{\rm e})$ can be written for $x_{\rm e}\gtrsim1$ as
\begin{equation}
\label{eq:powerspectrumsub}
{\cal{P}}_{\cal{R}}\left(x_{\rm{e}}\right)
=\frac{1}{4\pi^2}\frac{H_{\rm{e}}^4}{\dot\phi_{\rm{e}}^2}
\left[1+\frac{x_{\rm{e}}^2}{c_s^2}\right]\,.
\end{equation}
Evaluating the mean square at the approximate maximum,
$(x_*-x_{\rm{e}})\simeq1$,  Eq.~(\ref{eq:meansquarephi3}) for
$x_{\rm e}>1$ is approximately given by \footnote{This approximation breaks
down for $x_{\rm e}\leq1$, but the mean square value $\sigma_{\rm{\Psi}}$
converges very fast to the value given in 
Eq.~(\ref{eq:meansquarephi4}) as $x_{\rm e}$ increases over the value 1.}
\begin{eqnarray}
\label{eq:meansquarephi4}
\sigma_\Psi(m)
&\simeq&2{\cal{P}}^{1/2}_{\cal{R}}\left(x_{\rm{e}}\right)
\frac{x_{\rm{e}}}{x_*^2}\\
&=&2\mathcal{A}^{1/2}_{\cal{R}}\left(1+\frac{x^2_{\rm e}}{c_s^2}\right)^{1/2}
\frac{x_{\rm e}}{\left(1+x_{\rm e}\right)^{2}}\,,
\end{eqnarray}
where $\mathcal{A}^{1/2}_{\cal{R}}=(H_{\rm e}^2/\dot\phi_{\rm e})/2\pi$. 
Note that
the value of $\sigma_{\Psi}$ converges in the limit $x_{\rm e}\gg1$ to give
\be 
\sigma_{\Psi}
=2\mathcal{A}^{1/2}_{\cal{R}}/c_s=2\sqrt{3}\cala_\calr^{1/2}\,.  
\ee
This is a very important feature for the formation of PBHs inside the
horizon, as the number of mean deviations
$N_{\Psi}=\Psi_{\rm c}/\sigma_{\Psi}$ reaches a minimum value.

Once we have determined $\Psi_{\rm c}$ and $\sigma_{\Psi}$, we can write the
mass fraction of PBHs per logarithmic mass interval
\begin{equation}
\label{eq:Dmassfraction2}
m\frac{dF\left(m,\Psi_{c}\right)}{dm}
=-\sqrt{\frac{2}{\pi}}\frac{\Psi_{\rm{c}}}{\sigma_{\Psi}^2}
\frac{md\sigma_{\Psi}}{dm}
\exp\left(-\frac{\Psi^2_{\rm{c}}}{2\sigma_{\Psi}^2}\right)\,.
\end{equation}
The derivative of the mass with respect to the comoving scale is
\begin{equation}
\label{eq:dmass}
\frac{dm}{dk}
=-\frac{4\pi}{3}\left(\frac{a_{\rm{e}}}{k}\right)^3\frac{1}{k}\;
\frac{\left(2+3x_{\rm{e}}\right)}
{x_{\rm{e}}\left(1+\frac{1}{x_{\rm{e}}}\right)}
=-\frac{m}{k}\;\frac{2+3x_{\rm e}}{1+x_{\rm e}}\,.
\end{equation}
On the other hand the derivative of the mean square deviation with
respect to the comoving wavenumber yields
\begin{equation}
\label{eq:sigmaderivative}
\frac{d\sigma_{\rm{\Psi}}}{dk}
=\frac{d\sigma_{\rm{\Psi}}}{dx_{\rm{e}}}\frac{dx_{\rm{e}}}{dk}
=\frac{\sigma_{\rm{\Psi}}}{k}\frac{6x_{\rm e}^2-x+1}{\left(1+x_{\rm e}^2/c_s^2\right)
\left(1+x_{\rm e}\right)}\,.
\end{equation}
These two last expressions allow us to write the derivative of the
mean square deviation in the following form
\begin{equation}
\label{eq:sigmaderivative2}
\frac{d\sigma_{\rm\Psi}}{dm}
=\frac{d\sigma_{\rm\Psi}}{dx_{\rm e}}\frac{dx_{\rm e}}{dk}\frac{dk}{dm}
=-\frac{\sigma_{\rm{\Psi}}}{m}\;{\cal F}\left(x_{\rm e}\right)\,,
\end{equation}
where $\mathcal F(x_{\rm e})$ is defined as
\be
\label{calf_def}
{\cal{F}}\left(x_{\rm{e}}\right)
=\frac{6x_{\rm e}^2-x_{\rm e}+1}{\left(1+x_{\rm e}^2/c_s^2\right)\left(2+3x_{\rm e}\right)}\,.
\ee
Taking expression (\ref{eq:sigmaderivative2}) the mass spectrum can be
written as
\be
\label{eq:massfrac}
\frac{dn\left(m,\Psi_{c}\right)}{dm}
=\sqrt{\frac{2}{\pi}}\frac{\rho}{m^2}\mathcal{F}
\left(x_{\rm{e}}\right)N_{\Psi}\exp\left(-\frac{N_{\Psi}^2}{2}\right)\,,
\ee
where $N_{\Psi}=\Psi_{\rm c}/\sigma_{\Psi}$ is the number of mean deviations
of the Gaussian distribution of the metric potential $\Psi$,
and $\rho$ is the energy density of the background at formation time.

\section{Discussion and conclusion}
\label{disc_sec}

The discussion of the results becomes more transparent if we introduce
the quantity $(m^2/\rho)dn/dm$, the mass fraction of PBHs per
logarithmic interval of mass. Unlike the number density $dn/dm$, the
mass fraction is finite on all mass scales and therefore the total mass
of PBHs is finite as expected. {}From \eq{eq:massfrac} the mass fraction
$(m^2/\rho)dn/dm$ is
\be
\label{eq:massfracres}
\frac{m^2}{\rho}\frac{dn\left(m\right)}{dm}
=\sqrt{\frac{2}{\pi}}\mathcal{F}
\left(x_{\rm{e}}\right)N_{\Psi}\exp\left(-\frac{N_{\Psi}^2}{2}\right)\,,
\ee  
valid for 
${\cal{A}_R}^{3/2}m_{\rm e}\lesssim m <0.1 m_{\rm e}$,
where $m_{\rm e}$ is the horizon mass at the end of inflation,
$m_{\rm e}=(4\pi/3)\rho_{\rm e}H_{\rm e}^{-3}$. The lower bound is the
required limit according to cosmological linear theory. The upper bound on the
validity of Eq.~(\ref{eq:massfracres}), $m<0.1 m_{\rm e}$,
stems from the following requirements. The relation between $x_{\rm e}$ and 
the mass of the black hole inside
the horizon, $m$, is given by Eq.~(\ref{eq:PBHmass}), and we get
\be
\label{eq:massxe}
x_{\rm e}^2\left(x_{\rm e}+1\right)\simeq c_s^3\frac{m_{\rm e}}{m}\,.
\ee

Then the mass fraction inside the
horizon in Eq.~(\ref{eq:massfracres}) is valid for $x_{\rm e}>1$ or
$m<0.1m_{\rm e}$. For simplicity, in the following plots we have taken
$x_{\rm e}=c_s(m/m_{\rm e})^{-1/3}$. This amounts to neglecting the term
$\ln(1+1/x_{\rm e})$ when we take the logarithm of $m/m_{\rm e}$, and again the
approximation breaks down for $x_{\rm e}\lesssim1$.

\begin{figure}
\begin{center}
\includegraphics[angle=0,width=0.5\textwidth]{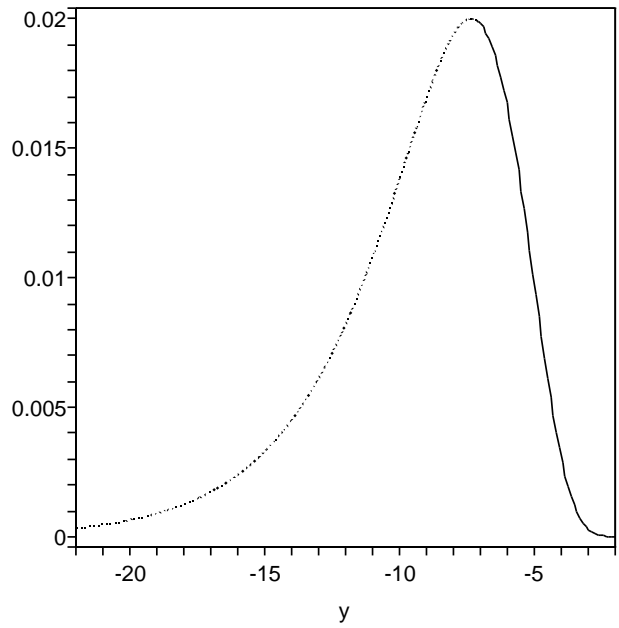}
\end{center}
\caption{\label{fig:1}
The mass fraction $(m^2/\rho)dn/dm$ versus $y=\ln(m/m_{\rm e})$ for
$\Psi_{\rm c}=0.5$ and ${\cal{A}_R}^{1/2}=0.09$. The dotted line correspond
to the extension of the mass fraction in Eq.~(\ref{eq:massfracres})
 beyond the lower bound $y\simeq7$.}
\end{figure}

\begin{figure}
\begin{center}
\includegraphics[angle=0,width=0.5\textwidth]
{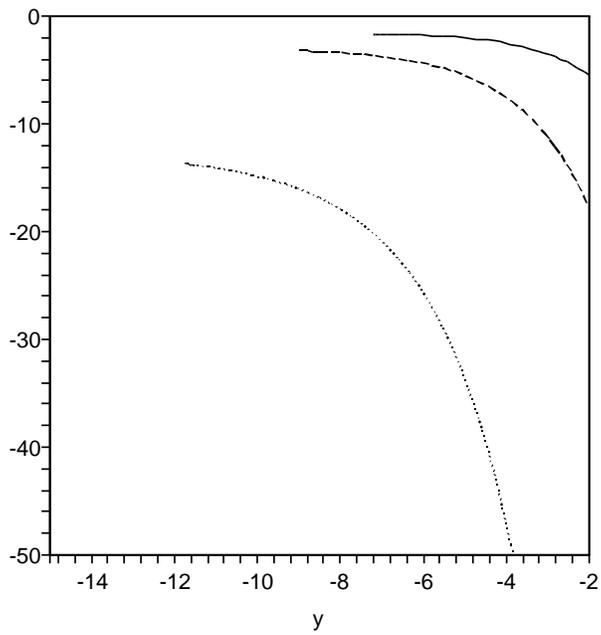}
\end{center}
\caption{\label{fig:2} 
$\log_{10}((m^2/\rho)dn/dm)$ versus $y=\ln(m/m_{\rm e})$ for 
$\Psi_{\rm c}=0.5$. The dotted line corresponds to the
amplitude ${\cal{A}_R}^{1/2}=0.02$, the dashed line to
${\cal{A}_R}^{1/2}=0.05$, and the full line to
${\cal{A}_R}^{1/2}=0.09$.}
\end{figure}

\begin{figure}
\begin{center}
\includegraphics[angle=0,width=0.5\textwidth]
{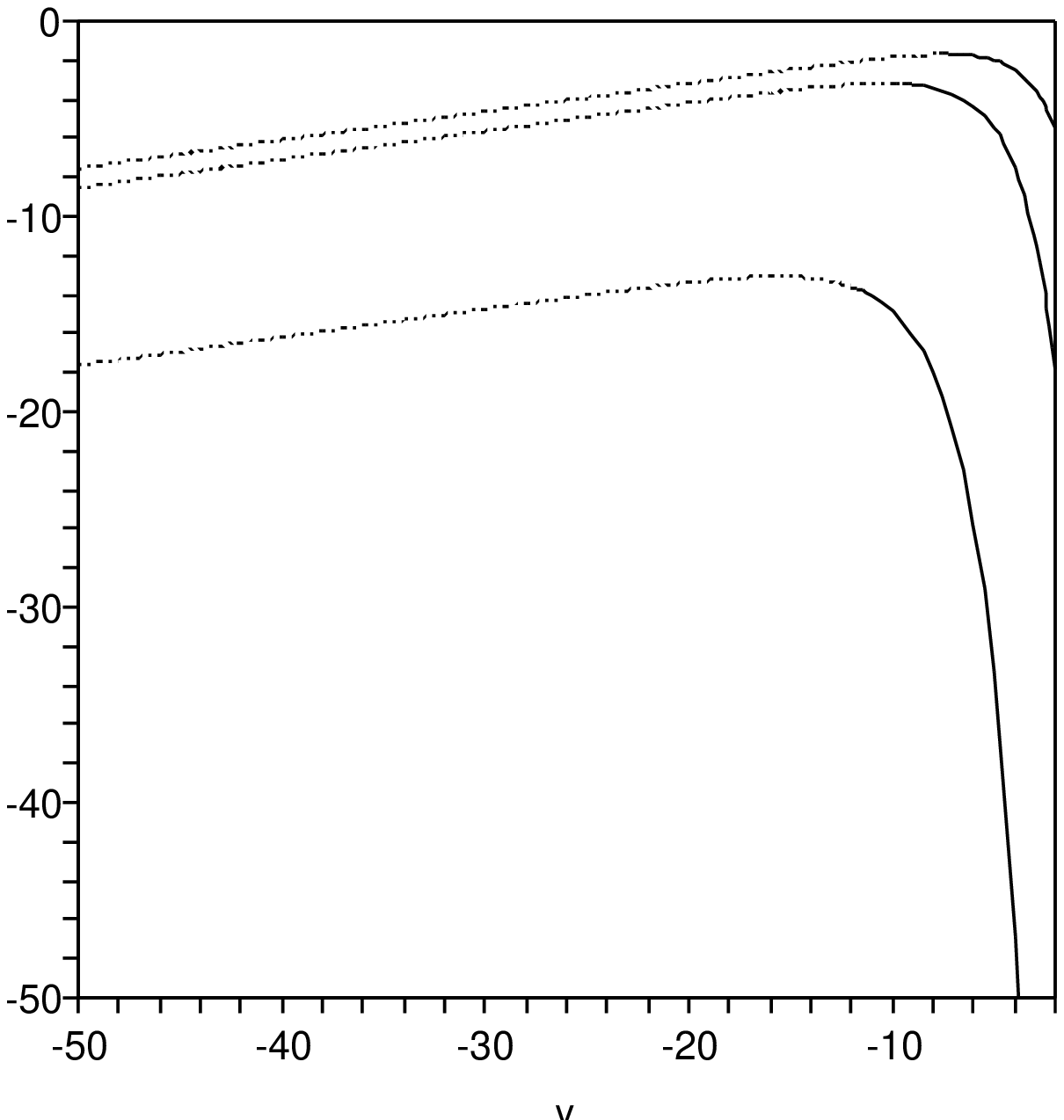}
\end{center}
\caption{\label{fig:3} 
$\log_{10}((m^2/\rho)dn/dm)$ versus $y=\ln(m/m_{\rm e})$ for
 $\Psi_{\rm c}=0.5$ and ${\cal{A}_R}^{1/2}=0.02$, $0.05$ and $0.09$. 
Here the dotted lines correspond to the extension of the mass fraction in
 Eq.~(\ref{eq:massfracres}) beyond the lower bound 
$m={\cal{A}_R}^{3/2}m_{\rm e}$.}
\end{figure}

Using the upper bound on the mean deviation of the density contrast at
horizon crossing in \cite{Leach:2000ea,Copeland:1998cs},
$\sigma_{HC}\lesssim0.04$, we can set an upper bound on the amplitude
of the fluctuations on sub-horizon scales
${\cal{A}_R}^{1/2}\simeq(9/4)\sigma_{HC}\lesssim 0.09$.
The following graphs are shown highlighting the regime in which our
calculation is valid. The mass fraction $(m^2/\rho) dn/dm$ inside the horizon
increases rapidly with the exponential factor as the number of mean deviations
$N_\Psi$ decrease. Hence the mass fraction rises sharply with $k$, 
as shown in Fig.~\ref{fig:1} with a solid line for a threshold value of
$\Psi_{\rm c}=0.5$ and an amplitude ${\cal{A}_R}^{1/2}=0.09$.
In Figs.~\ref{fig:2} and \ref{fig:3} we show the logarithm of the mass
fraction for a threshold value of $\Psi_{\rm c}=0.5$ and an amplitude
${\cal{A}_R}^{1/2}=0.09$, and also a more conservative amplitudes of
${\cal{A}_R}^{1/2}=0.05$ and $0.02$. 
The extension of the mass fraction in equation (\ref{eq:massfracres}) beyond
the limit $m={\cal{A}_R}^{3/2}m_{\rm e}$ is shown in dotted lines in 
Figs.~\ref{fig:1} and \ref{fig:3}, reaching a maximum and eventually
converging to zero.  

There are several things worth noting in Figs.~\ref{fig:2} and \ref{fig:3}: 
(i) The maximum is displaced to smaller scales as 
the number of mean deviations $N_{\Psi}$ increases.
(ii) As $N_\Psi$ increases the mass fraction maximum value falls
dramatically. The curve for ${\cal{A}_R}=0.01$, which is not shown in 
Fig.~\ref{fig:2}, has a maximum with an approximate value of $10^{-60}$.  
(iii) The maximum of the mass fraction and the lower bound nearly coincide 
for ${\cal{A}_R}^{1/2}=0.09$ on a mass scale of approximately
 $m\simeq7\times10^{-4}m_{\rm e}$. However, Fig.~\ref{fig:2} shows that the 
drift of the maximum into smaller scales is more rapid than the one of the
 current lower bound as ${\cal{A}_R}^{1/2}$ decreases its value. 


An estimate of the total mass fraction inside the horizon is
provided by integrating \eq{eq:massfracres} over the relevant range of masses. 
Considering a sharp cut-off at $m={\cal{A}_R}^{3/2}m_{\rm e}$, the total mass
fraction gives an upper limit of $4\times10^{-2}$ 
for a threshold value $\Psi_{\rm c}=0.5$ with an amplitude of 
${\cal{A}_R}=0.09$. If we extend the validity of Eq.~(\ref{eq:massfracres}) 
into the ultraviolet regime, the upper limit of the total mass fraction
increases to approximately $0.14$. The difference between these two estimates
of the total mass fraction increases considerably as ${\cal{A}_R}^{1/2}$ 
decreases. Therefore, the behaviour of the mass function in the ultraviolet 
regime needs further investigation. We postpone this issue and the detailed 
calculation of the observational effects of
sub-horizon PBH formation, in particular the implications for PBH
abundance calculations, to a future publication.

\acknowledgments

The authors are grateful to Bernard Carr and Anne Green for useful
discussions and comments.
The work of MS is supported in part by JSPS Grants-in-Aid
for Scientific Research (B)~17340075 and (S)~14102004.
The Lancaster group is supported by PPARC grants PPA/G/O/2002/00469
and PPA/V/S/2003/00104 and by EU grants HPRN-CT-2000-00152 and
MRTN-CT-2004-503369, and DHL is supported by PPARC grants
PPA/G/O/2002/00098 and PPA/S/2002/00272.




\begin{thebibliography}{99}

\bibitem{Liddle:2000cg}
A.~R.~Liddle and D.~H.~Lyth,
Cosmological inflation and large-scale structure, (CUP, Cambridge, 2000).

\bibitem{Carr:1974nx}
B.~J.~Carr and S.~W.~Hawking,
Mon.\ Not.\ Roy.\ Astron.\ Soc.\  {\bf 168}, 399 (1974).

\bibitem{Carr:1975qj}
B.~J.~Carr,
Astrophys.\ J.\  {\bf 201}, 1 (1975).

\bibitem{PS} W. H. Press and P. Schechter, 1974, Astrophys. J. {\bf 187}, 
	452 (1974).

\bibitem{Niemeyer:1999ak}
J.~C.~Niemeyer and K.~Jedamzik,
Phys.\ Rev.\ D {\bf 59}, 124013 (1999)
[arXiv:astro-ph/9901292].

\bibitem{Green:2004wb}
A.~M.~Green, A.~R.~Liddle, K.~A.~Malik and M.~Sasaki,
Phys.\ Rev.\ D {\bf 70}, 041502 (2004)
[arXiv:astro-ph/0403181].

\bibitem{Shibata:1999zs}
M.~Shibata and M.~Sasaki,
Phys.\ Rev.\ D {\bf 60}, 084002 (1999)
[arXiv:gr-qc/9905064].

\bibitem{Musco:2004ak}
I.~Musco, J.~C.~Miller and L.~Rezzolla,
Class.\ Quant.\ Grav.\  {\bf 22}, 1405 (2005)
[arXiv:gr-qc/0412063].

\bibitem{Kim:1996hr}
H.~I.~Kim and C.~H.~Lee,
Phys.\ Rev.\ D {\bf 54}, 6001 (1996).

\bibitem{Carr:1994ar}
B.~J.~Carr, J.~H.~Gilbert and J.~E.~Lidsey,
Phys.\ Rev.\ D {\bf 50}, 4853 (1994)
[arXiv:astro-ph/9405027].

\bibitem{Green:1997sz}
A.~M.~Green and A.~R.~Liddle,
Phys.\ Rev.\ D {\bf 56}, 6166 (1997)
[arXiv:astro-ph/9704251].

\bibitem{Leach:2000ea}
S.~M.~Leach, I.~J.~Grivell and A.~R.~Liddle,
Phys.\ Rev.\ D {\bf 62}, 043516 (2000)
[arXiv:astro-ph/0004296].

\bibitem{Carr:2005bd}
B.~J.~Carr,
eConf {\bf C041213}, 0204 (2004)
[arXiv:astro-ph/0504034].

\bibitem{Starobinsky82}
A.~A.~Starobinsky,
Phys.\ Lett.\ B {\bf 117}, 175 (1982).

\bibitem{Starobinsky86}
A.~A.~Starobinsky,
In De Vega, H.j. ( Ed.), Sanchez, N. ( Ed.): ``Field Theory,
                  Quantum Gravity and Strings'', 107-126, (1986).


\bibitem{decoherence}
M.~Schlosshauer,
Rev.\ Mod.\ Phys.\  {\bf 76}, 1267 (2004)
[arXiv:quant-ph/0312059].

\bibitem{Polarski:1995jg}
D.~Polarski and A.~A.~Starobinsky,
Class.\ Quant.\ Grav.\  {\bf 13}, 377 (1996)
[arXiv:gr-qc/9504030].

\bibitem{Kiefer:1998qe}
C.~Kiefer, D.~Polarski and A.~A.~Starobinsky,
Int.\ J.\ Mod.\ Phys.\ D {\bf 7}, 455 (1998)
[arXiv:gr-qc/9802003].

\bibitem{Bardeen:1980kt}
J.~M.~Bardeen,
Phys.\ Rev.\ D {\bf 22}, 1882 (1980).

\bibitem{Kodama:1985bj}
H.~Kodama and M.~Sasaki,
Prog.\ Theor.\ Phys.\ Suppl.\  {\bf 78}, 1 (1984).



\bibitem{Israel:1966rt}
W.~Israel,
Nuovo Cim.\ B {\bf 44S10}, 1 (1966)
[Erratum-ibid.\ B {\bf 48}, 463 (1967\ NUCIA,B44,1.1966)].


\bibitem{Deruelle:1995kd}
N.~Deruelle and V.~F.~Mukhanov,
Phys.\ Rev.\ D {\bf 52}, 5549 (1995)
[arXiv:gr-qc/9503050].


\bibitem{Martin:1997zd}
J.~Martin and D.~J.~Schwarz,
Phys.\ Rev.\ D {\bf 57}, 3302 (1998)
[arXiv:gr-qc/9704049].


\bibitem{Copeland:1998cs}
E.~J.~Copeland, A.~R.~Liddle, J.~E.~Lidsey and D.~Wands,
Gen.\ Rel.\ Grav.\  {\bf 30}, 1711 (1998)
[arXiv:gr-qc/9805073].







\end{thebibliography}
\end{document}